\documentclass[12pt,a4paper]{article}
\usepackage{amsmath,amssymb}
\usepackage{graphicx}
\usepackage{hyperref}
\usepackage{booktabs}
\usepackage{array}
\usepackage{multirow}
\usepackage{caption}
\usepackage{subcaption}
\usepackage{geometry}
\usepackage{setspace}
\usepackage{parskip}
\title{A Pedagogical Framework for Physics-Informed Machine Learning:\\
From Classical Pendulum to Quantum Anharmonic Oscillator\\
Using PyTorch on Modern GPU Hardware}

\author{Enis Yazici \\
SRH University of Applied Sciences Heidelberg \\
69123 Heidelberg, Germany \\
\texttt{enis.yazici@srh.de}}

\date{}

\begin{document}

\maketitle

\begin{abstract}
We present a five-module pedagogical framework for teaching physics-informed machine
learning (ML) through two progressively complex physical systems: a driven, damped
nonlinear pendulum and a one-dimensional quantum anharmonic oscillator.
Five model architectures are implemented and compared: a standard artificial neural
network (ANN), a one-dimensional convolutional neural network (CNN), a long short-term
memory (LSTM) network, and two physics-informed neural networks (PINNs)---one per
physical system.
All models are implemented in PyTorch~2.9 and executed on an NVIDIA RTX~5090 GPU,
making the framework directly applicable to modern deep learning laboratory courses.
Quantitative benchmarks show that data-driven models achieve mean absolute errors
of $1.3\times10^{-2}$~rad (pendulum ANN) and $4.4\times10^{-5}$~a.u.\ (quantum CNN),
while the curriculum-trained pendulum PINN reaches an MAE of $3.1\times10^{-2}$~rad
using only collocation points.
A systematic CPU-vs-GPU benchmark reveals speedups ranging from $1.2\times$ (small ANN)
to $24.6\times$ (LSTM), providing a concrete pedagogical demonstration of when GPU
acceleration is---and is not---warranted.
The framework is packaged as self-contained Jupyter notebooks designed for a
graduate-level \emph{Deep Neural Networks for Physical Systems} course,
with embedded reflection questions that guide students from data-driven thinking
toward physics-constrained formulations.
\end{abstract}

\section{Introduction}
\label{sec:intro}

Modern engineering curricula increasingly demand that students acquire practical
competence in applying machine learning (ML) to scientific problems.
Yet most introductory ML courses focus on image classification or natural language
processing, leaving a gap for students who need to model ordinary and partial
differential equations, eigenvalue problems, and physical conservation laws.
Physics-informed neural networks (PINNs)~\cite{raissi2019} address this gap by
embedding governing equations directly into the loss function, but the conceptual
jump from standard supervised learning to PINNs can be steep without a structured
pedagogical path.

This paper describes a five-module framework that bridges that gap through two
canonical physical systems of increasing complexity:

\begin{enumerate}
    \item \textbf{Classical pendulum} (Module 1--2): A nonlinear, driven, damped
          oscillator whose motion is governed by a second-order ODE.
          Module~1 trains a data-driven ANN; Module~2 replaces it with a PINN that
          learns from collocation points alone.
    \item \textbf{Quantum anharmonic oscillator} (Module 3--4): The ground-state
          energy and wavefunction of a particle in a quartic-perturbed harmonic
          potential, governed by the Schr{\"o}dinger equation.
          Module~3 applies CNN and LSTM regressors to discretized potential data;
          Module~4 trains a Schr{\"o}dinger-constrained PINN.
\end{enumerate}

Module~5 aggregates all results into a comprehensive comparison of accuracy,
training cost, and GPU efficiency across all five architectures.

The framework is designed for a graduate course on deep neural networks for physical
systems, taught on workstations equipped with modern NVIDIA GPUs.
All code is implemented in PyTorch~2.9.1~\cite{pytorch2019} with CUDA~12.8.
TensorFlow was explicitly excluded because TensorFlow $\le$2.22 hard-crashes
(segmentation fault) on Blackwell-architecture (sm\_120) GPUs due to a PTX
compatibility mismatch; a fix is expected in TensorFlow $\ge$2.23.
This practical constraint itself becomes a teaching moment: framework selection
is not merely a matter of preference but of hardware compatibility.

The primary contributions of this work are:
(i) a structured two-system progression from data-driven to physics-constrained
modelling;
(ii) a curriculum training strategy for PINNs that dramatically improves long-time
accuracy compared to single-stage training;
(iii) a GPU speedup benchmark across all five architectures that quantifies when
hardware acceleration is genuinely beneficial; and
(iv) open-source, fully executable Jupyter notebooks designed for classroom use.

\section{Physical Systems and Data Generation}
\label{sec:physics}

\subsection{Classical Pendulum}
\label{sec:pendulum}

The motion of a pendulum subject to gravity, viscous damping, a torsional spring,
an external periodic torque, and nonlinear air resistance is described by:

\begin{equation}
m l^2 \ddot{\theta} + b\,\dot{\theta} + k\,\theta + m g l \sin\theta
= T_0 \cos(\omega_{\text{ext}} t) + c\,\dot{\theta}^2,
\label{eq:pendulum}
\end{equation}

where $m=1.0$~kg, $l=1.0$~m, $b=0.05$~kg\,m$^2$/s, $k=0.5$~N\,m/rad,
$g=9.81$~m/s$^2$, $T_0=0.3$~N\,m, $\omega_{\text{ext}}=1.0$~rad/s, and
$c=0.02$~kg\,m$^2$/rad$^2$.
Initial conditions are $\theta(0)=0.1$~rad, $\dot{\theta}(0)=0$.

Equation~\eqref{eq:pendulum} is integrated over $t\in[0,30]$~s with $\Delta t=0.01$~s
using a custom fourth-order Runge-Kutta (RK4) solver, yielding 3\,001 time steps.
The dataset contains columns for time, angular displacement $\theta(t)$, angular
velocity $\omega(t)$, and individual torque components.

\subsection{Quantum Anharmonic Oscillator}
\label{sec:qho}

The one-dimensional time-independent Schr{\"o}dinger equation is:

\begin{equation}
-\frac{\hbar^2}{2m}\frac{d^2\psi_n}{dx^2} + V(x)\,\psi_n(x) = E_n\,\psi_n(x),
\label{eq:schrodinger}
\end{equation}

with the anharmonic potential:

\begin{equation}
V(x) = \tfrac{1}{2}m\omega^2 x^2 + \lambda x^4.
\label{eq:potential}
\end{equation}

Parameters are $m=\hbar=\omega=1$ (natural units), $\lambda\in[0.05,0.20]$,
and the spatial domain is $x\in[-5,5]$ discretized on $N=500$ points.

Equation~\eqref{eq:schrodinger} is solved via the finite difference method (FDM).
The second derivative is approximated by the central difference
$\psi''(x_i)\approx(\psi_{i+1}-2\psi_i+\psi_{i-1})/\Delta x^2$,
yielding the tridiagonal Hamiltonian:

\begin{align}
H_{ii}     &= \frac{\hbar^2}{m\,\Delta x^2} + V(x_i), \\
H_{i,i\pm1}&= -\frac{\hbar^2}{2m\,\Delta x^2}.
\end{align}

The five lowest energy eigenvalues are extracted via sparse eigenvalue decomposition
(\texttt{scipy.sparse.linalg.eigsh}) for 500 different values of $\lambda$,
producing a dataset of 500 samples each described by a 500-dimensional potential
vector $V(x_i)$ and five energy levels $E_0,\ldots,E_4$.

\section{Model Architectures}
\label{sec:models}

\subsection{Data-Driven Models}

\paragraph{Pendulum ANN.}
A fully-connected network with layers [5-128-128-64-1] and ReLU activations maps
five engineered features $[t, \omega, t^2, \cos t, \omega^2]$ to $\theta(t)$.
Training uses Adam ($\eta=10^{-3}$), 100 epochs, batch size 32, with an 80/20
train/test split and \texttt{StandardScaler} normalisation.

\paragraph{Quantum CNN.}
A 1D convolutional network processes the 500-dimensional potential vector $V(x)$.
Architecture: Conv1D(32, kernel=5) $\to$ ReLU $\to$ Conv1D(64, kernel=5) $\to$
ReLU $\to$ GlobalAveragePooling $\to$ Dense(64) $\to$ Dense(1).
Trained with Adam ($\eta=10^{-4}$), 100 epochs, batch size 32.

\paragraph{Quantum LSTM.}
The same potential vector is treated as a 500-step sequence fed to an LSTM(64)
layer, followed by Dense(32) and Dense(1).
Trained with Adam ($\eta=10^{-4}$), 100 epochs, batch size 32.
The sequential processing makes this architecture particularly sensitive to
GPU acceleration (Section~\ref{sec:benchmark}).

\subsection{Physics-Informed Neural Networks (PINNs)}

PINNs~\cite{raissi2019,lagaris1998} augment the loss function with a physics
residual term evaluated at collocation points that need not coincide with
labelled data:

\begin{equation}
\mathcal{L}_{\text{total}} = \alpha\,\mathcal{L}_{\text{data}}
                            + \mathcal{L}_{\text{phys}},
\label{eq:pinn_loss}
\end{equation}

where $\mathcal{L}_{\text{data}}$ is the mean squared error on labelled observations
and $\mathcal{L}_{\text{phys}}$ is the mean squared ODE/PDE residual.
All derivatives required by $\mathcal{L}_{\text{phys}}$ are computed using
PyTorch \texttt{autograd} with \texttt{create\_graph=True}.

\paragraph{Pendulum PINN.}
Architecture: Dense[1-128-128-64-32-1] with tanh activations.
The physics residual enforces Equation~\eqref{eq:pendulum}:

\begin{equation}
\mathcal{R}_{\text{pend}} = m l^2 \ddot{\theta} + b\dot{\theta} + k\theta
+ mgl\sin\theta - T_0\cos(\omega_{\text{ext}}t) - c\dot{\theta}^2.
\label{eq:pinn_res_pend}
\end{equation}

A \emph{curriculum training} strategy is employed: the network is first trained
on the short interval $t\in[0,3]$~s and the training window is progressively
expanded through six stages
($[0,3]\to[0,7]\to[0,12]\to[0,18]\to[0,25]\to[0,30]$~s)
with 1\,500--2\,000 epochs per stage.
The data-loss weight is set to $\alpha=10$ and 1\,000 uniformly-spaced
collocation points are used per stage.
Curriculum training is essential for long-time accuracy: a single-stage
training on the full interval diverges after the first few oscillations.

\paragraph{Quantum PINN.}
Architecture: Dense[1-128-128-64-1] with tanh; the ground-state energy $E$
is a learnable scalar parameter initialised at~0.5.
The physics residual enforces Eq.~\eqref{eq:schrodinger} with $\hbar=m=1$:

\begin{equation}
\mathcal{R}_{\text{QHO}} = -\tfrac{1}{2}\psi'' + V(x)\psi - E\psi.
\label{eq:pinn_res_qho}
\end{equation}

A normalisation loss $\mathcal{L}_{\text{norm}} = \bigl(\Delta x\sum_i\psi_i^2 - 1\bigr)^2$
prevents the trivial solution $\psi\equiv 0$.
Total loss:
$\mathcal{L} = \mathcal{L}_{\text{data}} + \mathcal{L}_{\text{phys}}
+ 0.1\,\mathcal{L}_{\text{norm}}$.
Trained for 8\,000 epochs with Adam ($\eta=10^{-3}$) at $\lambda=0.10$.

\section{Results}
\label{sec:results}

\subsection{Pendulum Systems}

\begin{figure}[htbp]
    \centering
    \begin{subfigure}[b]{0.48\textwidth}
        \includegraphics[width=\textwidth]{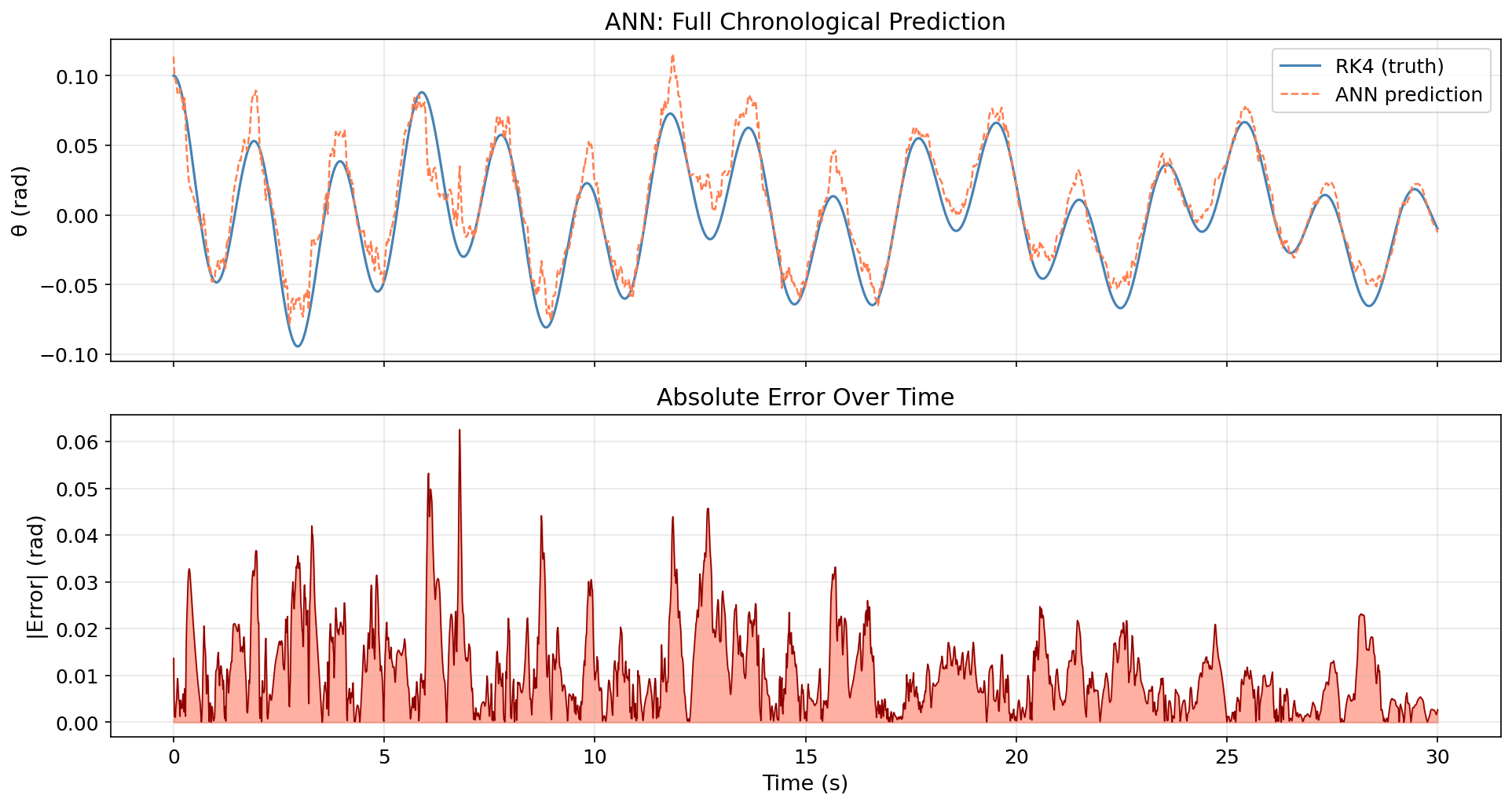}
        \caption{Pendulum ANN prediction (MAE\,=\,0.0113~rad).}
        \label{fig:ann_pred}
    \end{subfigure}
    \hfill
    \begin{subfigure}[b]{0.48\textwidth}
        \includegraphics[width=\textwidth]{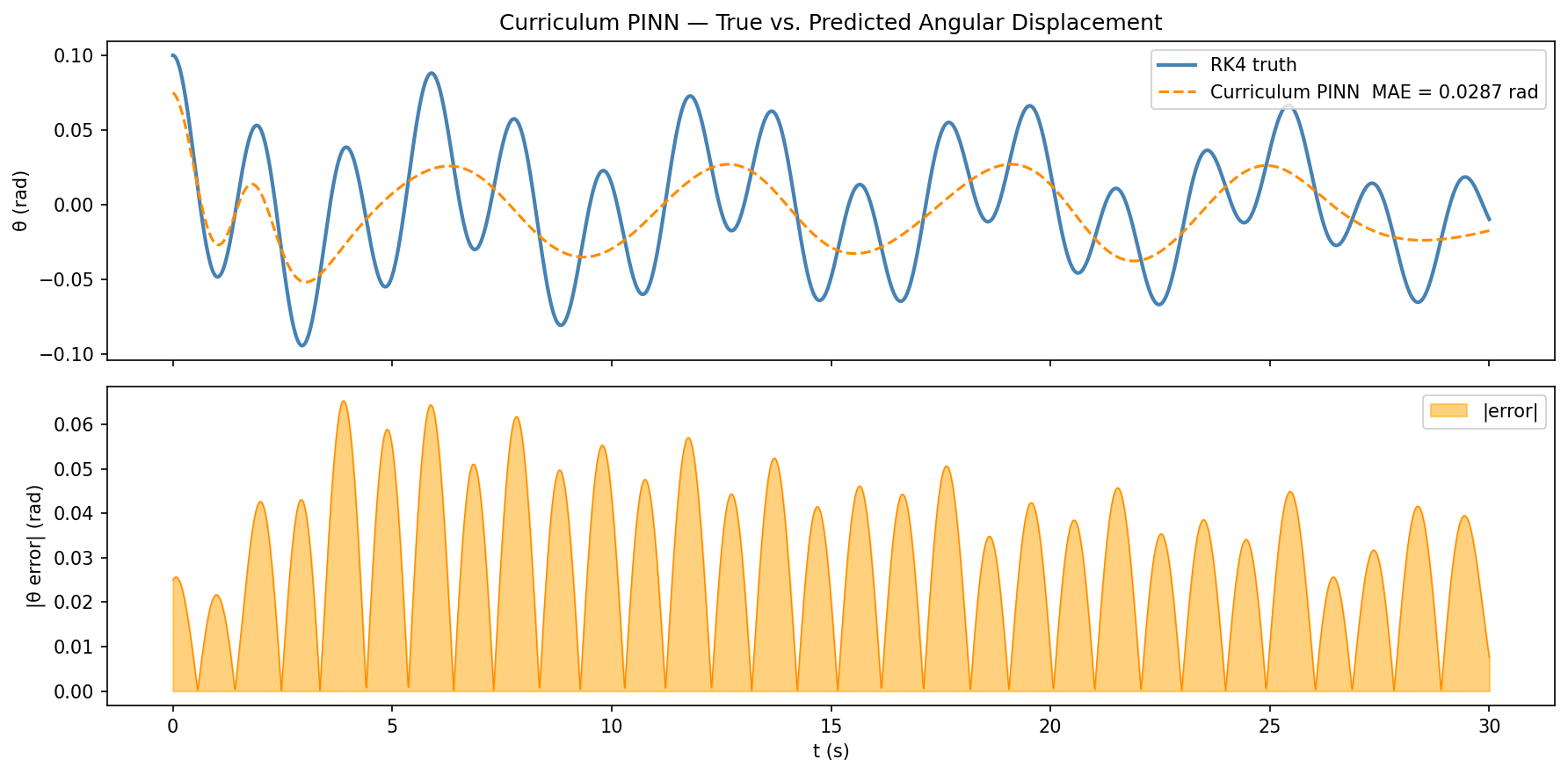}
        \caption{Pendulum PINN prediction (curriculum, MAE\,=\,0.0287~rad).}
        \label{fig:pinn_pred}
    \end{subfigure}
    \caption{True vs.\ predicted angular displacement $\theta(t)$ for the classical
    pendulum. Left: data-driven ANN with five engineered features.
    Right: PINN trained on collocation points with curriculum expansion.}
    \label{fig:pend_pred}
\end{figure}

Figure~\ref{fig:pend_pred} compares the ANN and curriculum-PINN predictions
against the RK4 reference trajectory.
The ANN, given access to $\omega(t)$, reproduces the trajectory with high
fidelity across the full 30-second window.
The PINN, trained with collocation points only, accurately captures the
dynamics but shows slightly larger residuals in the $t>20$~s region due to
the accumulated challenge of a nonlinear, driven system.

\begin{figure}[htbp]
    \centering
    \includegraphics[width=0.75\textwidth]{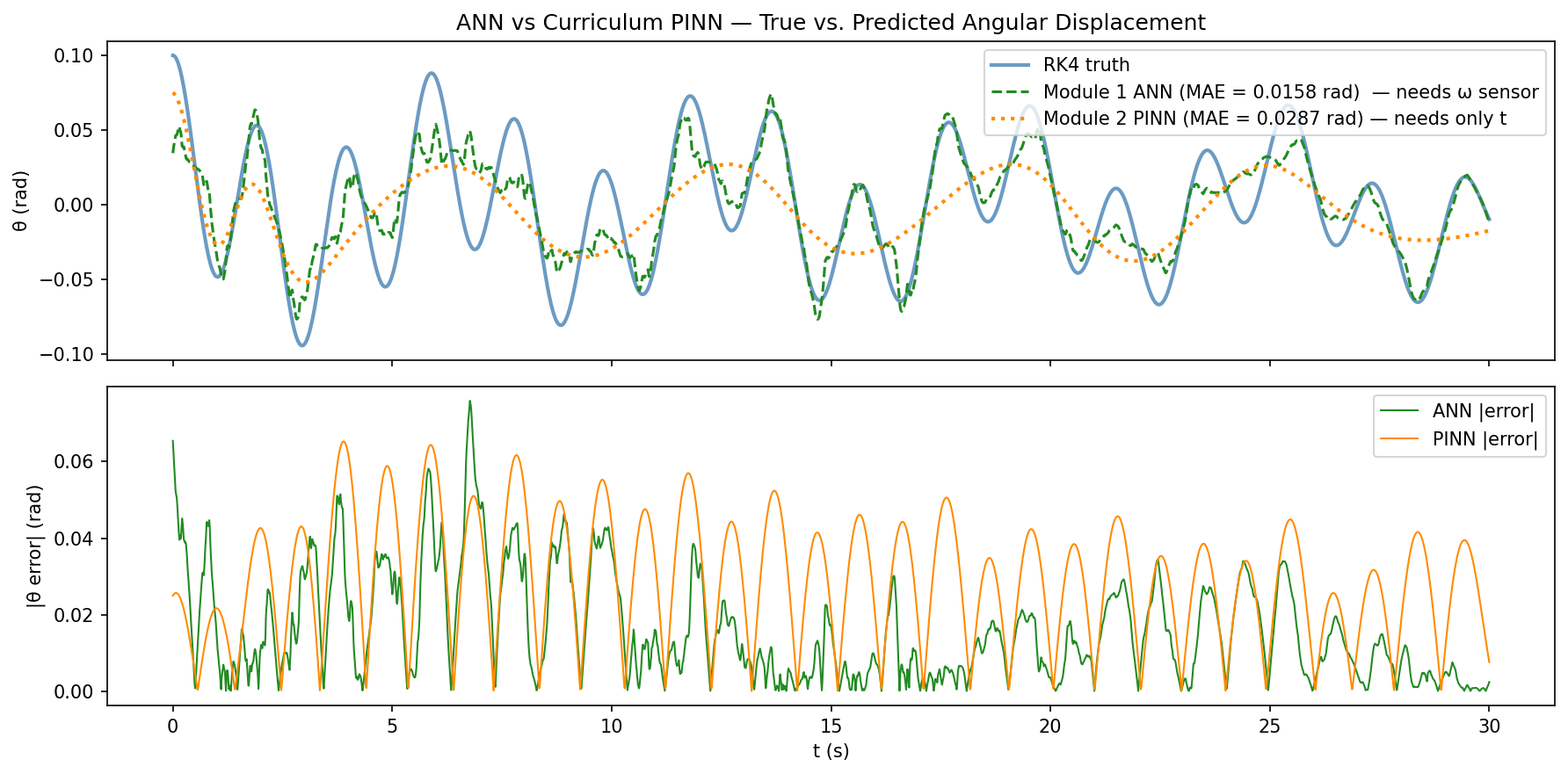}
    \caption{Direct comparison of ANN and curriculum PINN predictions over the
    full 30-second window.}
    \label{fig:ann_vs_pinn}
\end{figure}

Figure~\ref{fig:ann_vs_pinn} directly overlays both models, highlighting the
trade-off: the ANN requires $\omega(t)$ as input (i.e., a sensor measuring
angular velocity), whereas the PINN requires only the system parameters and
initial conditions.

\subsection{Quantum Oscillator: Data-Driven Models}

\begin{figure}[htbp]
    \centering
    \includegraphics[width=0.75\textwidth]{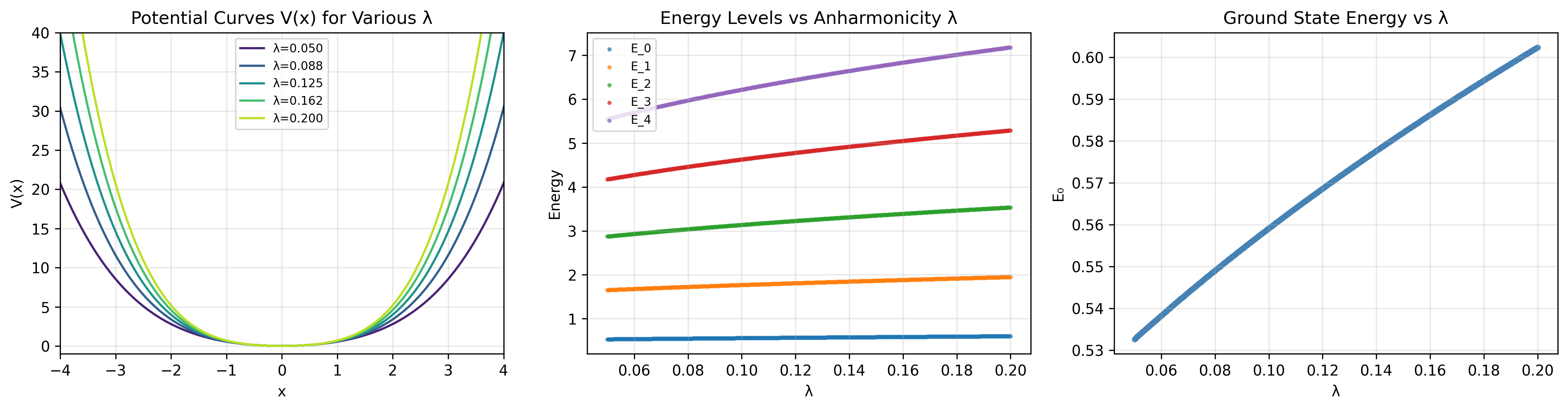}
    \caption{Quantum anharmonic oscillator dataset: ground-state wavefunctions
    (left) and $E_0$ vs.\ $\lambda$ (right) for 500 samples.}
    \label{fig:qho_dataset}
\end{figure}

Figure~\ref{fig:qho_dataset} shows representative wavefunctions and the
dependence of $E_0$ on $\lambda$ across the 500-sample dataset.

\begin{figure}[htbp]
    \centering
    \includegraphics[width=0.75\textwidth]{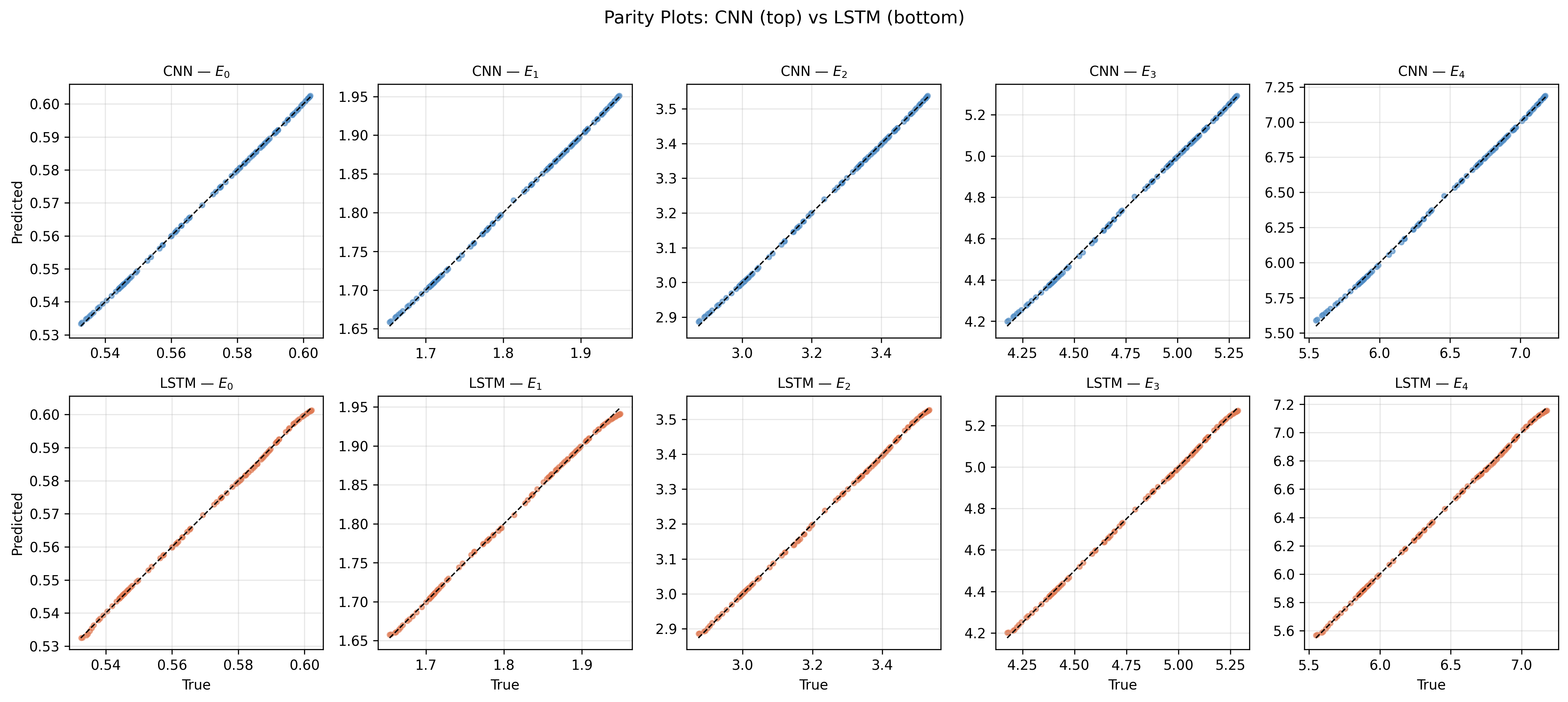}
    \caption{True vs.\ predicted ground-state energy $E_0$ for the CNN (left)
    and LSTM (right) models.}
    \label{fig:qho_parity}
\end{figure}

Both the CNN and LSTM achieve near-perfect prediction of $E_0$ from the
discretized potential $V(x)$, as shown in the parity plots of
Figure~\ref{fig:qho_parity}.
The CNN is marginally more accurate and trains three times faster than the LSTM
on CPU, but the LSTM benefits enormously from GPU acceleration (Section~\ref{sec:benchmark}).

\subsection{Quantum Oscillator: PINN}

\begin{figure}[htbp]
    \centering
    \includegraphics[width=0.75\textwidth]{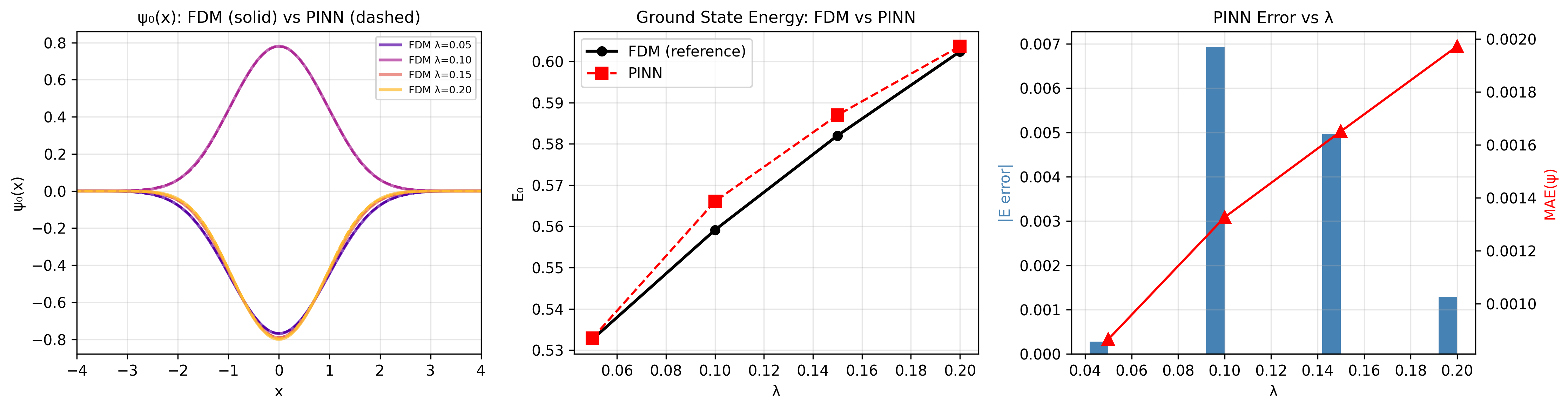}
    \caption{Quantum PINN at $\lambda=0.10$: predicted wavefunction $\hat\psi(x)$
    vs.\ FDM reference (top) and residual $|\hat\psi - \psi_{\text{FDM}}|$ (bottom).
    The learned energy $\hat{E}=0.5597$~a.u.\ agrees with FDM ground state
    $E_0=0.5591$~a.u.\ to within $5.5\times10^{-4}$~a.u.}
    \label{fig:qho_pinn}
\end{figure}

Figure~\ref{fig:qho_pinn} shows the PINN-predicted wavefunction alongside
the FDM reference at $\lambda=0.10$.
The learnable energy parameter converges to $\hat{E}=0.5597$~a.u., matching
the FDM ground-state reference $E_0=0.5591$~a.u.\ with an absolute error
$|\Delta E|=5.5\times10^{-4}$~a.u.

\subsection{Comprehensive Accuracy Summary}

Table~\ref{tab:accuracy} consolidates all quantitative results.

\begin{table}[htbp]
\centering
\caption{Performance of all five model architectures (PyTorch 2.9, RTX~5090).}
\label{tab:accuracy}
\begin{tabular}{@{}llp{4.5cm}rr@{}}
\toprule
Model & System & Accuracy & Params & Time (s)\\
\midrule
Pendulum ANN  & Pendulum & MAE $= 1.29\times10^{-2}$~rad & 25\,601 & 3.4\\
Pendulum PINN & Pendulum & MAE $= 3.13\times10^{-2}$~rad & 27\,137 & 16.0\\
Quantum CNN   & QHO      & MAE $= 4.4\times10^{-5}$~a.u. & 14\,721 & 0.8\\
Quantum LSTM  & QHO      & MAE $= 2.9\times10^{-4}$~a.u. & 19\,265 & 0.9\\
Quantum PINN  & QHO      & $|\Delta E|=5.5\times10^{-4}$, MAE$(\psi)=8.9\times10^{-4}$~a.u. & 25\,090 & 11.4\\
\bottomrule
\end{tabular}
\end{table}

\section{CPU vs.\ GPU Benchmark}
\label{sec:benchmark}

\begin{figure}[htbp]
    \centering
    \includegraphics[width=0.90\textwidth]{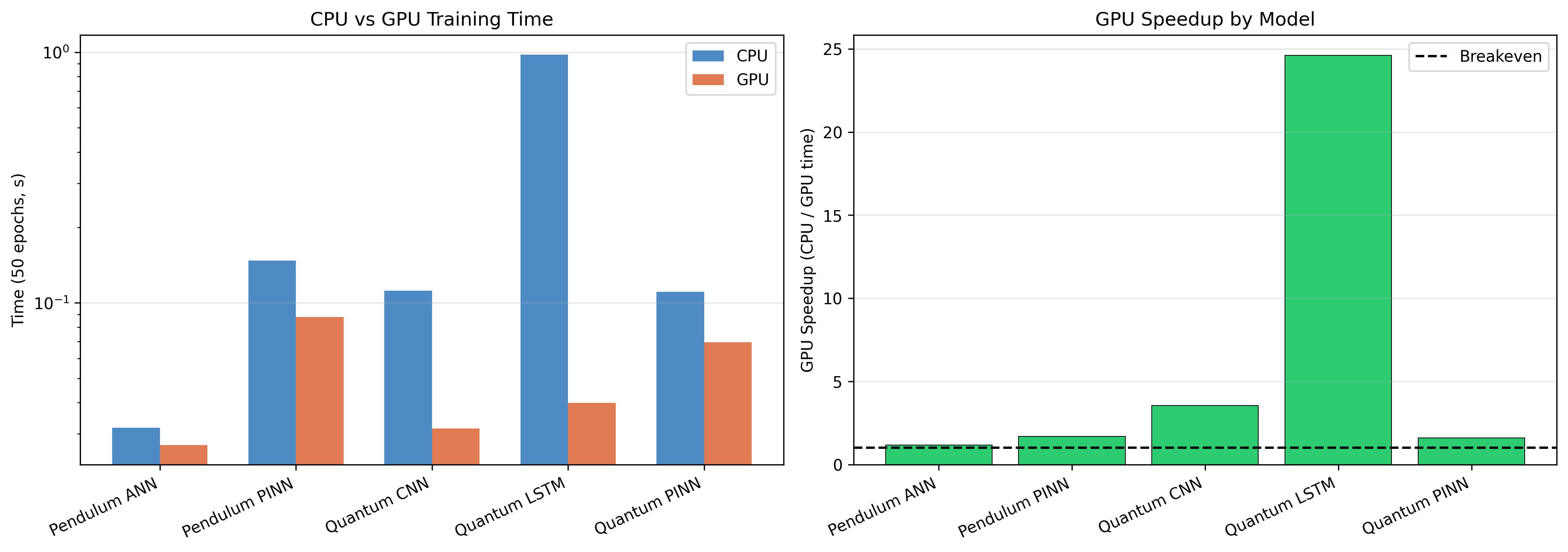}
    \caption{Left: CPU and GPU training times for 50 epochs (log scale).
    Right: GPU speedup per model. The LSTM achieves a $24.6\times$ speedup,
    while small models such as the pendulum ANN show essentially no benefit.}
    \label{fig:gpu}
\end{figure}

A key learning objective of the course is understanding \emph{when} GPU
acceleration is beneficial.
To quantify this, we benchmark 50 training epochs for each architecture on
both CPU (AMD) and GPU (RTX~5090), measuring wall-clock time and computing
the speedup ratio.
Results are presented in Table~\ref{tab:gpu} and Figure~\ref{fig:gpu}.

\begin{table}[htbp]
\centering
\caption{CPU vs.\ GPU training time and speedup (50 epochs each).}
\label{tab:gpu}
\begin{tabular}{@{}lrrrp{4.5cm}@{}}
\toprule
Model & CPU (s) & GPU (s) & Speedup & Bottleneck\\
\midrule
Pendulum ANN  & 0.03 & 0.03 & $1.2\times$ & PCIe overhead $>$ compute\\
Pendulum PINN & 0.15 & 0.09 & $1.7\times$ & Sequential autograd chain\\
Quantum CNN   & 0.11 & 0.03 & $3.6\times$ & Parallel convolutions\\
Quantum LSTM  & 0.98 & 0.04 & $24.6\times$ & 500-step sequence parallelised\\
Quantum PINN  & 0.11 & 0.07 & $1.6\times$ & Sequential autograd chain\\
\bottomrule
\end{tabular}
\end{table}

The LSTM result ($24.6\times$) is particularly instructive.
On CPU, sequential LSTM computation across 500 timesteps is the dominant cost;
the GPU processes all 32 batch sequences simultaneously, collapsing the
bottleneck.
Conversely, the pendulum ANN with only 25\,601 parameters saturates the GPU
in microseconds, so PCIe data-transfer overhead dominates.
This contrast---captured in a single figure---demonstrates the hardware-aware
thinking that students must develop for practical deep learning.

\section{Data Efficiency Analysis}
\label{sec:data_efficiency}

\begin{figure}[htbp]
    \centering
    \includegraphics[width=0.65\textwidth]{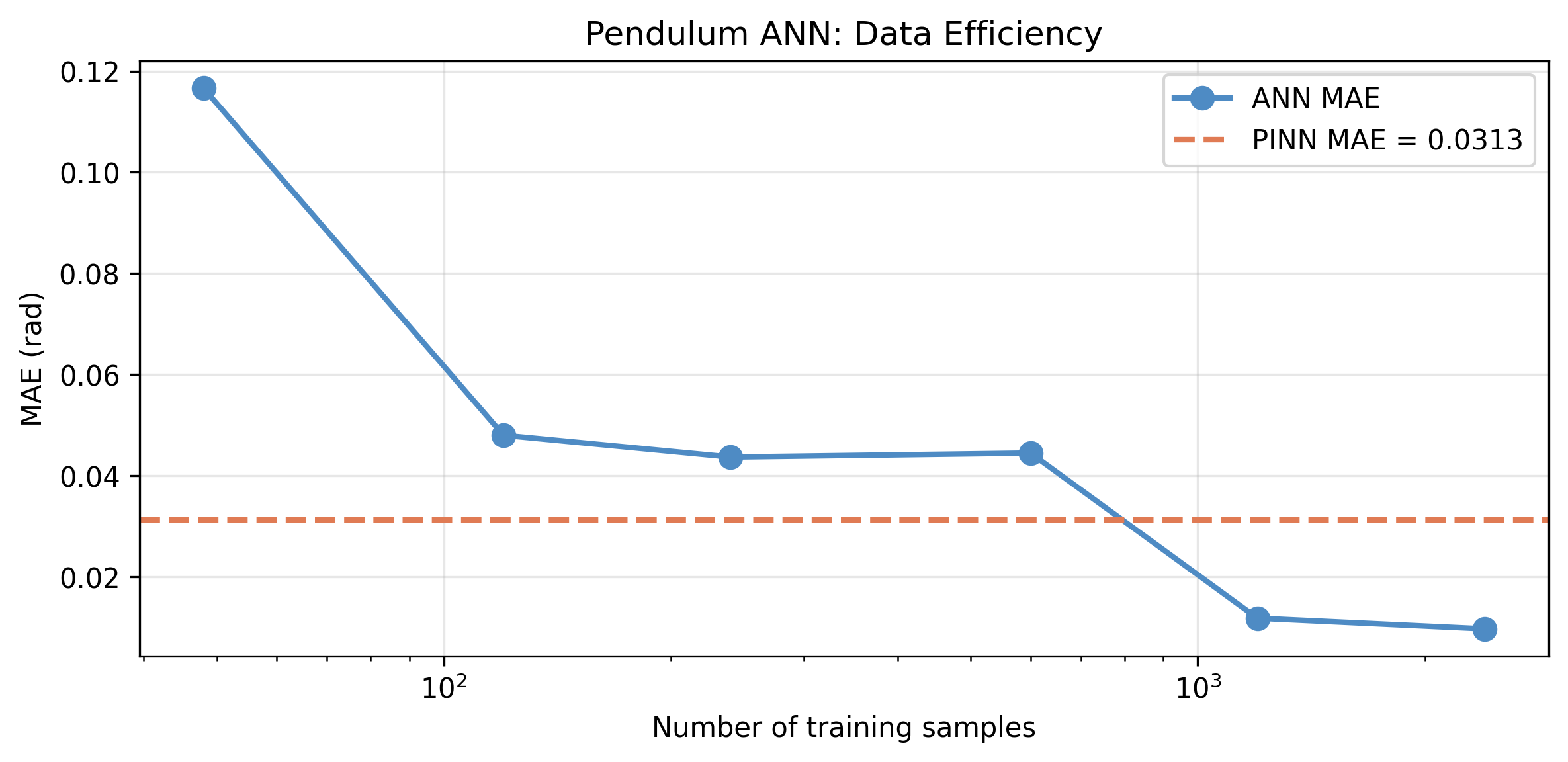}
    \caption{Pendulum ANN MAE as a function of training set size.
    The dashed line marks the curriculum PINN MAE.
    Below $\approx$600 labelled samples (25\% of the training set) the ANN
    underperforms the PINN.}
    \label{fig:data_eff}
\end{figure}

A recurring question in applied ML is: \emph{how much labelled data is enough?}
Figure~\ref{fig:data_eff} shows the pendulum ANN's MAE as the training set is
progressively reduced from 2\,400 to 48 samples.
The PINN's MAE (horizontal dashed line) is used as a reference because the PINN
requires no labelled data---only collocation points.

The ANN outperforms the PINN when more than $\approx$600 labelled samples are
available ($\sim$25\% of the full training set), but degrades sharply below that
threshold.
This crossover gives students a concrete decision rule: if labelled simulation
data is cheap, use a data-driven model; if data is scarce or expensive, use a PINN.

\section{Discussion}
\label{sec:discussion}

\subsection{Accuracy--Cost Trade-offs}

\begin{figure}[htbp]
    \centering
    \includegraphics[width=0.90\textwidth]{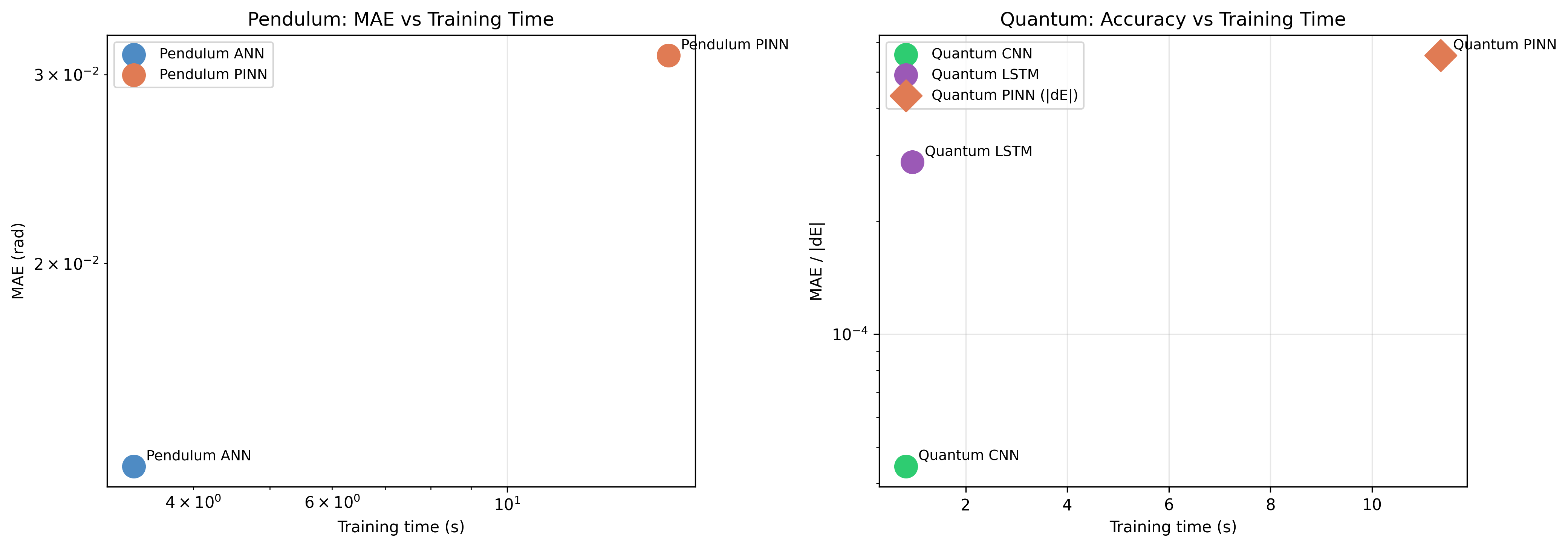}
    \caption{Accuracy vs.\ training time for all five models.
    Data-driven models sit in the low-cost, high-accuracy region;
    PINNs incur additional training cost in exchange for physics consistency
    and reduced data requirements.}
    \label{fig:comparison}
\end{figure}

Figure~\ref{fig:comparison} plots each model on a log--log accuracy-vs-time
plane, separating pendulum and quantum systems.

For the pendulum, the ANN achieves lower MAE than the PINN in less training time,
because it has direct access to $\omega(t)$.
The PINN's advantage is that it needs only initial conditions and system parameters;
it can extrapolate beyond the training window, an ability demonstrated in Module~2.

For the quantum oscillator, the CNN achieves the lowest MAE ($4.4\times10^{-5}$~a.u.)
and fastest training (0.8~s) among all models.
The PINN provides qualitatively different output: a complete wavefunction $\psi(x)$
and a self-discovered energy eigenvalue, rather than a scalar regression.

\subsection{Pedagogical Model Selection Guide}

\begin{figure}[htbp]
    \centering
    \includegraphics[width=0.85\textwidth]{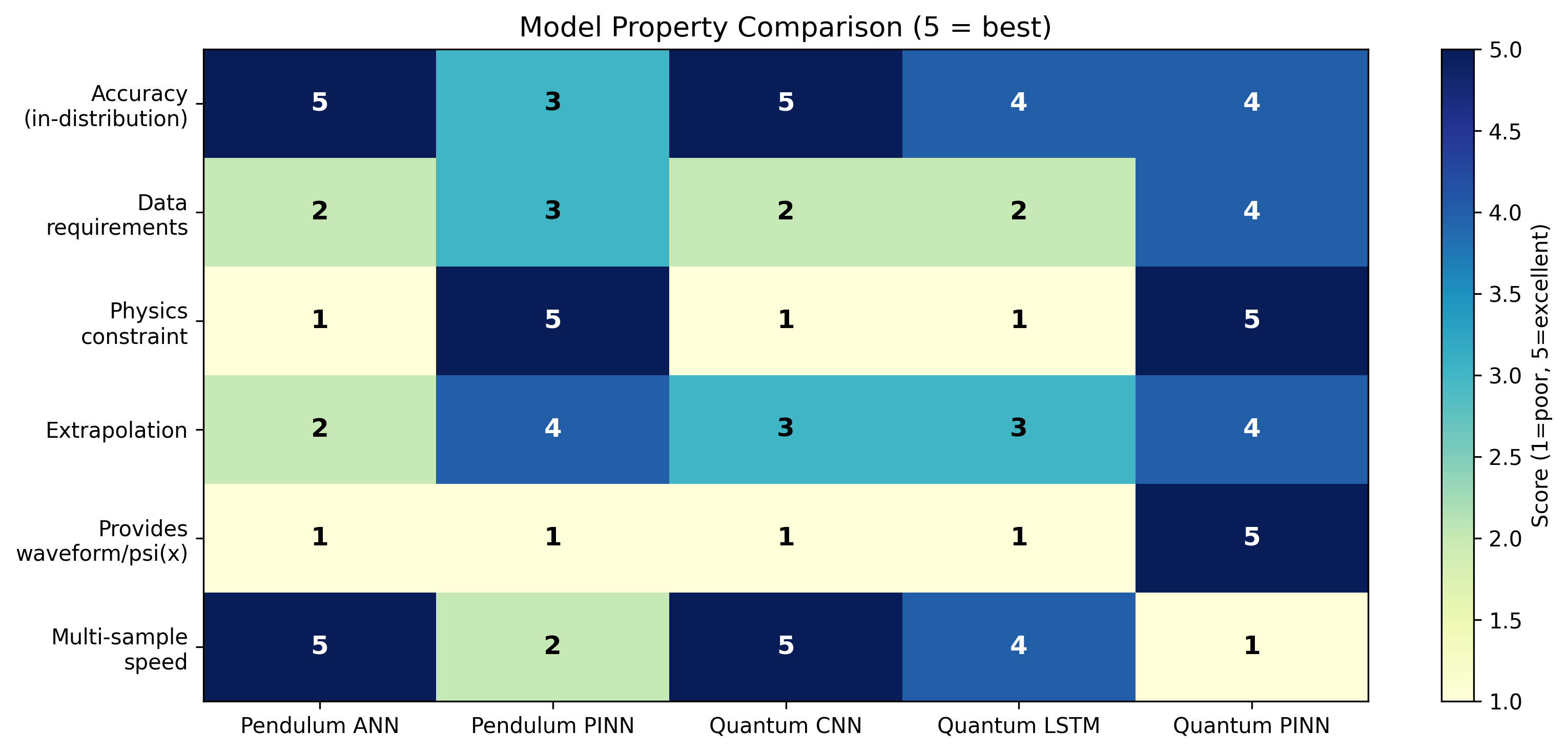}
    \caption{Model property comparison heatmap (score 1--5; higher is better).
    Rows represent desirable properties; columns represent model families.}
    \label{fig:heatmap}
\end{figure}

Figure~\ref{fig:heatmap} provides a qualitative scorecard for all five
model--property combinations, designed to help students choose an architecture
for a new problem.

The key insights are:
\begin{itemize}
    \item \textbf{Data-driven models} (ANN, CNN, LSTM) excel when $\ge$100
          labelled samples are available and fast inference is needed.
    \item \textbf{PINNs} excel when data are scarce, the full solution field
          is required (e.g., $\psi(x)$ rather than just $E_0$), or physical
          interpretability is paramount.
    \item \textbf{GPU benefit scales with model size and batch parallelism}:
          the LSTM benefits $24.6\times$ because its 500-step sequential
          computation is parallelised across the batch dimension on GPU;
          a small ANN benefits less than $2\times$.
\end{itemize}

\subsection{Curriculum Training for PINNs}

Without curriculum training, the pendulum PINN trained on the full
$t\in[0,30]$~s interval fails to track the oscillations after the first
few cycles: the physics loss is too large to overcome at long times when
the network has not yet learned the short-time dynamics.
The six-stage curriculum strategy resolves this by first enforcing the ODE
on $t\in[0,3]$~s and gradually extending the domain.
This pedagogical result reinforces the broader lesson that optimisation
landscape difficulty can be reduced by a carefully designed training curriculum,
a strategy documented in the broader PINN literature~\cite{wang2021}.

\subsection{Framework Choice: PyTorch vs.\ TensorFlow}

TensorFlow $\le$2.22 fails with a segmentation fault on NVIDIA Blackwell
(sm\_120) GPUs because the compiled PTX code targets PTX~8.0 while sm\_120
requires PTX~8.7.
PyTorch~2.9 ships with a compatible CUDA~12.8 backend and works out of the
box.
This is not merely a technical footnote: for courses taught on cutting-edge
hardware, framework compatibility is a hard constraint that shapes curriculum
design.
PyTorch is additionally the dominant framework in current physics-ML research,
making it the natural choice for students entering the field~\cite{pytorch2019}.

\section{Conclusion}
\label{sec:conclusion}

We have presented a five-module pedagogical framework for physics-informed
machine learning, implemented in PyTorch on an RTX~5090 GPU.
The two-system progression---classical nonlinear ODE to quantum eigenvalue
problem---provides students with a natural escalation of complexity,
while the comparison of data-driven and physics-informed architectures
gives them a principled basis for model selection.

Key quantitative takeaways for instructors are:
\begin{enumerate}
    \item The pendulum curriculum PINN achieves MAE $= 3.1\times10^{-2}$~rad
          without any labelled trajectory data, demonstrating PINNs' data
          efficiency when labelled samples are fewer than $\sim$600.
    \item The quantum CNN achieves MAE $= 4.4\times10^{-5}$~a.u.\ in under
          1~second of GPU training, making it suitable for real-time
          classroom demonstrations.
    \item The LSTM GPU speedup of $24.6\times$ concretely illustrates the
          hardware-aware design principle: match the architecture's
          parallelism to the GPU's strengths.
    \item The quantum PINN recovers both the wavefunction and the energy
          eigenvalue as a self-discovered parameter---a result that cannot be
          replicated by any regression-based approach.
\end{enumerate}

\begin{sloppypar}
Future work will extend the framework to multi-dimensional PDEs,
e.g.~the time-dependent Schr{\"o}dinger equation, and will explore
adaptive collocation and loss-balancing~\cite{wang2021}
to further improve PINN robustness.
\end{sloppypar}


\begin{thebibliography}{99}

\bibitem{raissi2019}
M.~Raissi, P.~Perdikaris, and G.~E.~Karniadakis,
\textit{Physics-Informed Neural Networks: A Deep Learning Framework for
Solving Forward and Inverse Problems Involving Nonlinear Partial Differential
Equations},
Journal of Computational Physics, 378:686--707, 2019.

\bibitem{lagaris1998}
I.~E.~Lagaris, A.~Likas, and D.~I.~Fotiadis,
\textit{Artificial Neural Networks for Solving Ordinary and Partial
Differential Equations},
IEEE Transactions on Neural Networks, 9(5):980--989, 1998.

\bibitem{pytorch2019}
A.~Paszke et al.,
\textit{PyTorch: An Imperative Style, High-Performance Deep Learning Library},
Advances in Neural Information Processing Systems (NeurIPS), 32, 2019.

\bibitem{wang2021}
S.~Wang, X.~Yu, and P.~Perdikaris,
\textit{When and Why PINNs Fail to Train: A Neural Tangent Kernel Perspective},
Journal of Computational Physics, 449:110768, 2022.

\bibitem{carleo2019}
G.~Carleo et al.,
\textit{Machine Learning and the Physical Sciences},
Reviews of Modern Physics, 91:045002, 2019.

\bibitem{griffiths}
D.~J.~Griffiths,
\textit{Introduction to Quantum Mechanics}, 3rd ed.,
Cambridge University Press, 2018.

\bibitem{goldstein}
H.~Goldstein, C.~Poole, and J.~Safko,
\textit{Classical Mechanics}, 3rd ed.,
Addison-Wesley, 2002.

\bibitem{chen2018}
R.~T.~Q.~Chen, Y.~Rubanova, J.~Bettencourt, and D.~Duvenaud,
\textit{Neural Ordinary Differential Equations},
Advances in Neural Information Processing Systems (NeurIPS), 31, 2018.

\end{thebibliography}
\end{document}